\documentclass[twocolumn,showpacs,superscriptaddress,preprintnumbers,amsmath,
amssymb,prb]{revtex4}

\usepackage{graphicx}% Include figure files
\usepackage{dcolumn}% Align table columns on decimal point
\usepackage{bm}% bold math
\usepackage{hyperref}% Hypertext references

\begin{document}

\title{Magnetic structure and orbital ordering in BaCoO$_3$ from
first-principles calculations}

\author{V. Pardo}
 \email{vpardo@usc.es}
\affiliation{
Departamento de F\'{\i}sica Aplicada, Facultad de F\'{\i}sica, Universidad
de Santiago de Compostela, E-15782 Campus Sur s/n, Santiago de Compostela,
Spain
}
\affiliation{
Institute for Materials Chemistry, Vienna University of Technology, 
Getreidemarkt 9/165, A-1060 Vienna, Austria}
\affiliation{
Instituto de Investigaciones Tecnol\'ogicas, Universidad de Santiago de
Compostela, E-15782, Santiago de Compostela, Spain
}

\author{P. Blaha}
\affiliation{
Institute for Materials Chemistry, Vienna University of Technology, 
Getreidemarkt 9/165, A-1060 Vienna, Austria}

\author{M. Iglesias}
\affiliation{
Departamento de F\'{\i}sica Aplicada, Facultad de F\'{\i}sica, Universidad
de Santiago de Compostela, E-15782 Campus Sur s/n, Santiago de Compostela,
Spain
}
\affiliation{
Instituto de Investigaciones Tecnol\'ogicas, Universidad de Santiago de
Compostela, E-15782, Santiago de Compostela, Spain
}

\author{K. Schwarz}
\affiliation{
Institute for Materials Chemistry, Vienna University of Technology, 
Getreidemarkt 9/165, A-1060 Vienna, Austria}

\author{D. Baldomir}
\affiliation{
Departamento de F\'{\i}sica Aplicada, Facultad de F\'{\i}sica, Universidad
de Santiago de Compostela, E-15782 Campus Sur s/n, Santiago de Compostela,
Spain
}
\affiliation{
Instituto de Investigaciones Tecnol\'ogicas, Universidad de Santiago de
Compostela, E-15782, Santiago de Compostela, Spain
}

\author{J.E. Arias}
\affiliation{
Instituto de Investigaciones Tecnol\'ogicas, Universidad de Santiago de
Compostela, E-15782, Santiago de Compostela, Spain
}

\date{\today}

\begin{abstract}

Ab initio calculations using the APW+lo method as implemented in the WIEN2k
code have been used to describe the electronic structure of the 
quasi-one-dimensional  system BaCoO$_3$. Both,
GGA and LDA+U approximations were employed to study different orbital and
magnetic orderings. GGA predicts a metallic ground state whereas  LDA+U 
calculations yield an insulating and ferromagnetic ground state (in a 
low-spin state) with an alternating
orbital ordering along the Co-Co chains, consistent 
with the available experimental data. 

\end{abstract}


\maketitle

\section{\label{intro}Introduction}

In the past few years, the main interest in Co oxides has been their 
applications 
in solid state fuel cells.\cite{thermo} Furthermore, some mixed valence Co
compounds showed interesting colossal magnetoresistance (CMR) 
effects.\cite{cmr}
Very recently, a renewed interest in Co oxides started
after the discovery of the first Co-based superconductor.\cite{nature}
Due to these two phenomena, CMR and superconductivity, Co oxides are of
great interest from a physical point of view.
Most of the interesting physical properties in transition metal
oxides lie in the interplay between electronic structure and magnetic
properties. 
New experimental techniques allow to investigate
the coupling between magnetic and orbital degrees of freedom, which plays
a major role in this type of materials.\cite{lacoo3,japan,OO}

BaCoO$_{3}$ is a transition metal oxide with interesting properties, which have
been studied in  recent years.\cite{yamaura,felser,ssc} In its
highly anisotropic structure, face-sharing 
CoO$_{6}$ octahedra form chains that are likely to produce one-dimensional 
effects in its magnetic properties. The
hexagonal structure, however, may lead to magnetic frustration when 
the Co atoms perpendicular to the chains (on a triangular arrangement)
should have an antiferromagnetic coupling.

The experimental studies on BaCoO$_3$ have established that the Co ions are
in a low-spin state (S=1/2),\cite{yamaura} as to be expected for a Co$^{4+}$ 
ion, with the configuration 
t$_{2g}^{5}$e$_{g}^{0}$ in the case of octahedral symmetry. 
In the temperature range from 70 to 300 K, the material is found to be
semiconducting, with
conduction occuring through n-type carriers.\cite{yamaura} The origin of 
the gap is 
not yet resolved.\cite{felser} Possible reasons are a Mott-Hubbard-like 
transition, an Anderson localization due to some structural disorder 
or a ``Peierls''-dimerization of the Co chains, but none of them has been
confirmed experimentally.

First measurements reported a N\'eel temperature T$_{N}$=8 K,
\cite{takeda} but recent experiments reveal a more complex magnetic
structure\cite{yamaura} beyond the simple antiferromagnetic behavior. 
Paramagnetism is found above 250 K, where
a coupling constant of J/k$_B$=10 K is estimated by fitting to a 1D
Heisenberg model. For temperatures between 70
and 250 K, ferromagnetic coupling is predominant, whereas for T$<$70 K
antiferromagnetic couplings become more important. 
Unfortunately no neutron diffraction data are available and
the magnetic order at very low temperatures is not known. Also the
relative strengths of intra-chain and inter-chain
magnetic couplings are still uncertain.

In this paper, we try to shed some light into the electronic structure of
BaCoO$_{3}$ by means of ab initio full-potential augmented plane wave plus
local orbitals (APW+lo) calculations\cite{sjo} carried
out with the WIEN2k code.\cite{wien,wien2k} We study some possible 
magnetic and orbital orderings (section \ref{struct})
and investigate the electronic structure 
from our ab initio calculations (section \ref{calcul}). 
In section \ref{ggasec} we present the results from the standard
GGA calculations and in section \ref{ldausec} we try to incorporate
electron correlation effects by means of the LDA+U approximation.

\section{\label{struct}Structure and magnetic orderings}

\begin{figure}
\includegraphics{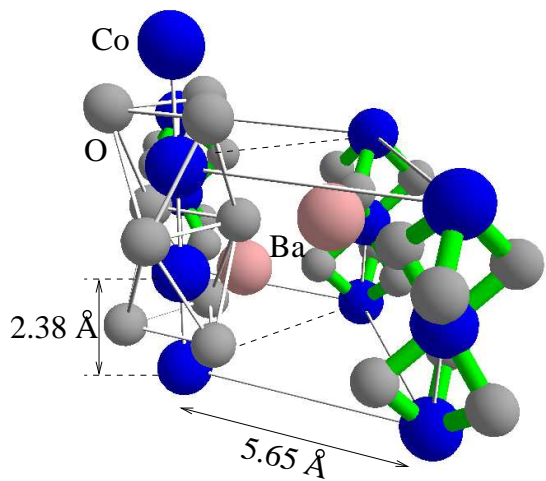}
\caption{\label{figstruct}(Color online) Hexagonal structure of BaCoO$_3$.  
Observe the chains of
face-sharing CoO$_6$ octahedra and the triangular arrangement of these chains
forming a hexagonal structure. 
}
\end{figure}

BaCoO$_{3}$ crystallizes in the hexagonal-2H pseudo-perovskite structure
with space group P6$_3$/mmc, and
lattice parameters a=5.645 \AA \  and c=4.752 \AA.\cite{struct} 
The position of the oxygen atoms has a free structural parameter whose
value was taken from Ref. \onlinecite{ssc} where it has been obtained from LAPW
calculations.

Figure \ref{figstruct} shows that BaCoO$_3$ forms chains
of face-sharing CoO$_6$ octahedra with a much shorter
($\simeq$ 2.38 \AA) Co-Co distance along the c axis than between the chains
($\simeq$ 5.65 \AA), giving the structure a strong 1D character. The
Co chains are organized in a triangular manner, giving rise to magnetic
frustration when antiferromagnetism occurs in the plane perpendicular to
the Co chains and assuming collinear magnetism. The Ba ion acts merely as 
an electron donor.

\begin{figure}
\includegraphics{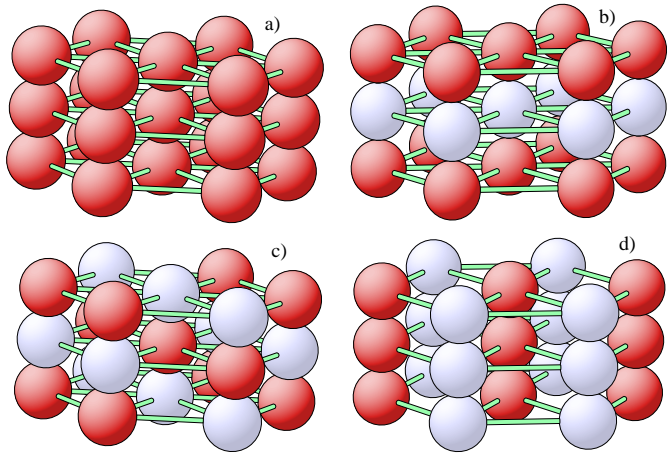}
\caption{\label{figmagn}(Color online) Different magnetic configurations 
of BaCoO$_3$
studied. Only Co atoms are shown, dark (light) spheres indicate spin up
(down) character.
a) FM; b) A-type AF (FM in plane, AF between planes); 
c) AF-type I (AF along c); d) AF-type II (FM along c). 
}
\end{figure}

In Fig. \ref{figmagn} we show the different magnetic orderings
studied in this work. Besides a ferromagnetic (FM) ordering (Fig. 
\ref{figmagn}a) also an ``A-type''
antiferromagnetic (AF) order is possible, where FM hexagonal planes are
coupled antiferromagnetically (Fig. \ref{figmagn}b).  The
triangular geometry of the Co chains leads to magnetic frustration of any
possible collinear AF long-range order in the hexagonal plane (perpendicular to
the chains).\cite{magnfrustr} Nevertheless, we have studied two partly AF
arrangements, where each Co has 2 FM and 4 AF
neighbors within the hexagonal plane. These planes can be coupled
antiferromagnetically (AF-type I, (Fig. \ref{figmagn}c)) or ferromagnetically
(AF-type II, (Fig. \ref{figmagn}d)) along the c axis.

We study these different configurations mainly in order to explore
the relative
strength of inter-chain with respect to the intra-chain magnetic
interactions. The quasi-one-dimensional structure of the compound would
imply that
the intra-chain magnetic coupling is most important, since
the corresponding Co-Co distance is more than two times smaller than the 
interchain Co-Co distance.

\section{\label{calcul}Computational details}

All calculations in this paper were carried out with the WIEN2k
package.\cite{wien,wien2k} This uses the full-potential APW+lo method\cite{sjo}
that makes no
shape approximation to the potential or density. Local orbitals (Co
3s, 3p, 3d, Ba 4d, 5s, 5p, O 2s and 2p) were added
to improve the flexibility of the basis set and to describe semicore states. 
The values of the atomic sphere
radii ($R_{mt}$) were chosen as: 1.70 a.u. for Co, 2.0 a.u. for Ba and 1.55 
a.u. for O.

We calculated the properties of the system within both the generalized 
gradient approximation (GGA) using the
PBE scheme\cite{gga} and the LDA+U approximation.\cite{sic2,anisimov,mazin} 
In the
latter, the DFT (orbital independent) part of the potential was treated by
the same GGA scheme. For 
GGA, we explored the energy vs. magnetic moment in the range
S=0 to S=1/2 using the fixed spin moment (FSM) method\cite{fsm} that fixes 
the total magnetic
moment in the unit cell. We used a plane wave cutoff described by
R$_{mt}$K$_{max}$=7 and 72 k-points in the irreducible Brillouin zone (IBZ)
(1000
k-points in the whole Brillouin zone). Convergence was checked up to 
R$_{mt}$K$_{max}$=8
and 150 k-points in the IBZ. For the larger AF unit cells,
equivalent k-meshes were employed.

The use of the LDA+U method is especially suited for this moderately 
correlated transition metal oxide, since it allows to introduce an
on-site Coulomb repulsion term, characterized by a Hubbard U. In addition, 
the introduction of an
orbital dependent potential allows to study orbital
orderings, which turn out to be an important issue in this class of compounds.

\section{GGA calculations}\label{ggasec}

\subsection{FSM curve}

In order to investigate the magnetic properties of the present compound, we 
analyzed the energy of the FM system by varying the magnetic moment of
the cell close to the experimentally found low-spin (LS) state. The
unit cell contains 2 Co ions, and thus the expected total magnetic moment for a
perfect LS state of Co would be 2 $\mu_B$. For this purpose we used the
FSM method within the GGA approximation for the
exchange-correlation potential. This method allows to calculate the total
energy as a function of the fixed total magnetization in the unit 
cell.\cite{fsm}

\begin{figure}
\includegraphics{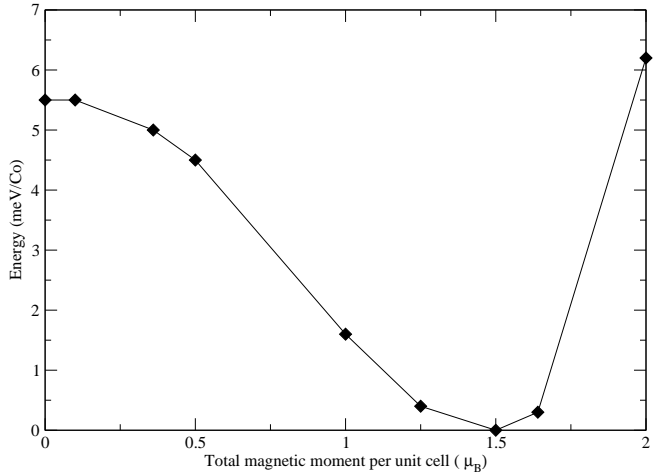}
\caption{\label{figfsm}Energy vs magnetic moment by FSM calculation using
GGA. The minimum is
near 1.5 $\mu_B$ while the perfect LS state (2 $\mu_B$) is comparable in 
energy to the nonmagnetic state. The line is just a guide for the eye.
}
\end{figure}

The results are shown in Fig. \ref{figfsm}, where the minimum of the FSM curve
is located near (at around 1.5 $\mu_B$ per unit cell) but definitely
below the LS state. Only about 0.5 $\mu_B$ reside inside each Co
sphere, but the rest of the magnetic moment resides inside
the O spheres and the interstitial region.
The LS state would imply
a half-metallic electronic structure, i.e. a gap in the
spin-up, but a metallic spin-down DOS, in contrast to the
experimentally observed semiconducting behavior, which cannot be obtained 
within GGA for the FM case.

\begin{figure}
\includegraphics{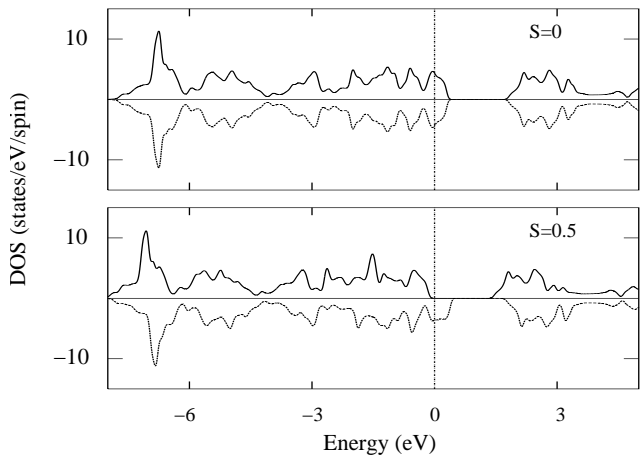}
\caption{\label{dosfsm}Total DOS curves within GGA for fixed values of the
total magnetic moment per unit cell of 0 and 2$\mu_B$, with S=0 and S=0.5,
respectively. 
}
\end{figure}

Figure \ref{dosfsm} shows how the DOS curves change when 
the total magnetic moment is varied from 0 to 2$\mu_B$. 
Half-metallicity is reached when 
the magnetic moment per Co site is set to an integer value. The 
spin-down compared to the spin-up DOS is approximately shifted rigidly.

These results roughly agree with LMTO-ASA calculations,
\cite{felser} which find the ground state to be a FM
low-spin state. They find, however, a much higher stabilization for that 
state with
respect to the nonmagnetic case, which is probably due to the atomic
sphere approximation (ASA).

\subsection{Antiferromagnetic calculations}

The nonsymmorphic symmetry operations involving a translation of (0,0,1/2) 
along the Co chains cause
degenerate bands that appear near the Fermi level.
To get an insulating behavior those degenerate bands 
must be split by breaking this symmetry, e.g. with an antiferromagnetic
order.
Within GGA calculations, the A-type AF order is unstable and converges
to a nonmagnetic solution. Among the cases studied here, the only AF 
structure that 
converges to a solution different from a nonmagnetic case is the 
AF-type II structure (see Fig. \ref{figmagn}d). This ordering assumes
a partial AF coupling of FM Co chains. The
in-plane coupling is FM along the b axis, leading to 2 FM but 4
AF nearest neighbors.
This solution has a fairly small magnetic moment on the Co sites, of just 
0.11$\mu_B$ per Co sphere. Although the Co chains
are quite far appart, 
we obtain a much smaller moment with this partly AF coupling than 
the completely FM aligned case (0.48
$\mu_B$ per Co sphere). Apparently the inter-chain coupling is much
stronger than one would intuitively assume according to the 
quasi-one-dimensional structure of the material.
The energy of the AF-type II structure is higher than the FM
case, which is the most stable solution within GGA, and is even higher than
a nonmagnetic solution (see Table \ref{energytab}).
We see that the energy differences involved are fairly small.

\begin{table}
\caption{\label{energytab}Relative energies (with respect to the FM
ground state) of the different magnetic
configurations studied with GGA. The value of the magnetic moment is for
the Co atom within its atomic sphere.}
\begin{tabular}{ccc}
\hline
\hline
& Energy (meV/Co) & Magnetic Moment ($\mu_B$)\\
\hline
Ferromagnetic & 0 & 0.48\\
Nonmagnetic & 5.5 & 0.0\\
AF-type II& 5.8 & 0.11\\
\hline
\end{tabular}
\end{table}

\subsection{Peierls distortion}

Another symmetry breaking, which could open a gap around the Fermi level, to
yield the experimentally observed semiconducting state, would be  
a dimerization of the Co atoms along the chains. 
In order to simulate the Peierls distortion, we displace one of the Co atoms
in the Co chains along the c axis towards the other Co to form dimers, i.e
one Co is displaced from the equilibrium position halfway between the
other type of Co atoms, which are at the corners of the unit cell (see
Fig. \ref{figstruct}).
For this constrained Co dispalacement, the positions of the oxygen and 
barium atoms were relaxed by minimizing the forces acting on them.

Even at very high distortions, we did not observe a semiconducting state,
contrary to the results from
LMTO-ASA calculations\cite{felser} (probably because these authors did not 
fully relax
the O and Ba atoms). We take oxygen relaxation into account and observe a 
flattening of the bands around the Fermi level, but the hybridization makes 
the bands broad enough to overlap causing a metallic behavior. 
We observe a decrease in the total magnetic
moment per unit cell from 1.5 $\mu_B$ in the symmetric case to 1.3 $\mu_B$
in the most distorted case studied. Since the corresponding energies are
high, we can conclude that in this system a Peierls distortion is not likely to
occur. A small Co distortion of 0.07 \AA \  from the equilibrium raises
the energy by 71 meV/Co but a larger distortion (of 0.32 \AA) by 1 eV/Co.
In the earlier bandstructure study,\cite{ssc} it has been shown that no Fermi
surface nesting is present.

\section{LDA+U calculations}\label{ldausec}

In previous sections we have shown that it is not possible to find an
insulating behavior within GGA, not even by symmetry breaking.
The LDA+U method within the so-called ``fully localized limit" \cite{mazin}
is a scheme that approximately allows to treat the electron-electron
interactions that occur in transition metal oxides. 
This method is known to improve over GGA in these moderately correlated 
transition metal oxides and to
predict better values of the magnetic moments and gaps.\cite{anisimov}

The method needs two input values, the on-site Coulomb repulsion U and the 
exchange energy J. The value of J was taken as the average
energy shift between spin-up and spin-down bands of the GGA spin-polarized
calculation (approximately 0.5 eV). For this type of
compounds the value of U is known to be in the range of several eV up to 
about 8 eV and was assumed 
to be 5 eV. We checked how the results depend on U and found that for
U=2.5 eV, we basically recover the GGA solution. For bigger values 
of U (6 eV and 7 eV), the FM solution remains
stable with a total magnetic moment of 1.00 $\mu_B$ per Co ion, but the energy
difference between the FM and AF solutions gets
gradually reduced as U increases. Also, the magnetic moments get more localized
inside the atomic spheres as U increases, as expected.

\begin{table}
\caption{\label{magnldau} Main
results for the different magnetic configurations studied within LDA+U
(U=5eV). Orbital order labelled ``alt. c" means that the hole is alternating 
between
d$_{x^{2}-y^{2}}$ and d$_{xy}$ orbitals in Co atoms
along the Co chain, ``alt. ab" means that it is alternating within the
hexagonal plane. The magnetic moments M correspond to the amount of the
moment that resides inside the atomic spheres
specified.}
\begin{tabular}{cccccc}
\hline
\hline
Magnetic & Orbital & Energy  & M(Co1) & M(Co2) & Insulator \\
order    & order   & (meV/Co)& ($\mu_B$) &($\mu_B$)&   \\
\hline
& alt. c & 0  & 0.846 & 0.850 & Yes \\
FM & d$_{x^{2}-y^{2}}$ & 71 & 0.698 & 0.698 & No\\
& d$_{xy}$ & 76 & 0.705 & 0.705 & No \\
\hline
 & alt. c & 7 & 0.688 & -0.689 & Yes\\
A-type AF & d$_{x^{2}-y^{2}}$ & 52 & 0.768 & -0.767 & Yes\\
& d$_{xy}$ & 54 & 0.770 & -0.770 & Yes \\
\hline
nonmagnetic & none & 58 & 0 & 0 & No\\
\hline
& alt. c & 12  & 0.811 & -0.818 & Yes \\
AF-type I & d$_{x^{2}-y^{2}}$ & 58  & 0.897& -0.897 & Yes\\
& d$_{xy}$ & 57 & 0.900 & -0.900 & Yes \\
\hline
& alt. ab & 86  & 0.651 & -0.674 & No \\
AF-type II & d$_{x^{2}-y^{2}}$ & 93 & 0.683 & -0.683 & No\\
& d$_{xy}$ & 93 & 0.682 & -0.682 & No \\
\hline
\end{tabular}
\end{table}

\subsection{Possible magnetic and orbital orderings}

We investigated the different magnetic
configurations shown in Fig. \ref{figmagn} using the LDA+U method. 
In this case, a
magnetic solution is possible for all structures, with strongly localized 
spin moments on the Co sites
and negligible contributions from the other spheres. The
main results are summarized in Table \ref{magnldau}.

\begin{figure}
\includegraphics{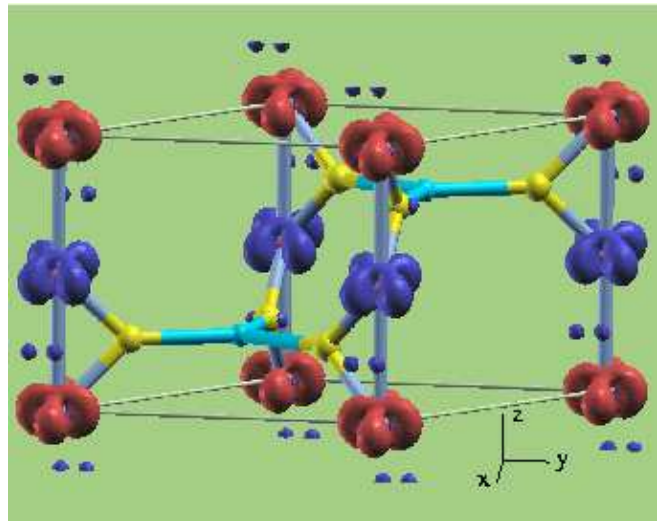}
\caption{\label{rho}(Color online) Spin density plot (isosurface at 0.2
e/\AA$^3$) for the alternating orbital ordered A-type AF
configuration. The spin moment is located in d$_{x^2-y^2}$ and d$_{xy}$
orbitals alternating along the
Co chains, (the orbitals form angles of 45 degrees with each other). 
The FM
orbital-ordered solution would be equivalent but with all Co in equal
color (same spin).
}
\end{figure}

Co$^{4+}$ in a (nearly) octahedral environment would be a 
t$_{2g}^{5}$e$_{g}^{0}$
system with one hole in the spin-down $t_{2g}$ states. In the present
hexagonal crystal structure, these are the d$_{z^2}$, d$_{xy}$ and 
d$_{x^2-y^2}$ orbitals, whereas
the  d$_{xz}$ and d$_{yz}$ orbitals form the e$_g$ manifold
pointing towards the O ligands. It turns out that within LDA+U the
hole can be localized predominatly in the  d$_{x^2-y^2}$ and/or  d$_{xy}$
orbitals, but never in the d$_{z^2}$ state, which is always fully occupied.
From Table \ref{magnldau}, we see that the
energy difference between solutions with a d$_{x^2-y^2}$ or d$_{xy}$ hole is
very  small, but both are much higher than the solution with an  alternating 
orbital ordering (i.e. Co atoms along the Co chain have alternating
d$_{x^2-y^2}$ and d$_{xy}$ holes).
This energy gain is much larger than the differences between different magnetic
order (FM vs. A-type AF or AF-type I or even a nonmagnetic solution).
Therefore,
magnetic configurations play a minor role in this quasi-one-dimensional
system. This finding is consistent with the results in Ref.
\onlinecite{mostovoi} where it was shown that this type of orbital ordering 
is caused by
superexchange interactions in a frustrated Jahn-Teller system with 
metal-oxygen-metal bonds of 90$^\circ$. In BaCoO$_3$, this angle is 86.9
$^\circ$ and thus spin and orbital degrees of freedom are decoupled, making the
spin exchange much weaker than the orbital one. 

The A-type AF structure is the most stable antiferromagnetic configuration
studied here.
It is the only one that can lead to a collinear long-range AF ordering
(along the Co chains). As mentioned above, the energy
difference to the FM ground state decreases with increasing U.

In the AF-type II case we restricted the Co atoms to
be equivalent
within the chain, and allowed orbital order only between the chains. Even 
this case of inter-chain orbital ordering is energetically 
favourable (though not as much as within the Co chain). Once again this is a
signal  for sizable inter-chain interactions, which are not expected to be
so large in a quasi-one-dimensional system.
 
For the FM case, the introduction of an alternating orbital
ordering along the chains 
removes the symmetry that makes the two types of Co atoms along c
equivalent and thus allows to
produce an energy gap around the Fermi level, leading to an insulating
behavior. Such an insulating state can also be reached when this
symmetry is broken by an AF ordering along the chain.
In the case of AF-type II, the intra-chain symmetry is not broken
and thus a metallic state is found. Just as in GGA, the FM solution is
the most stable in LDA+U calculations. In the latter case, the total
magnetic moment within the unit cell (which now contains 4 Co atoms) is 4.00
$\mu_B$ for the alternating orbital ordering (which corresponds to
the LS state), while it is only 3.9 $\mu_B$ for
the other orbital orderings (with metallic character). 

Figure \ref{rho} shows the spin density of the alternating orbital
ordered state of the A-type AF structure. 
The FM case would look similar, but the colors
(indicating spin-up and spin-down) would be equal for every Co atom. 
The spin-density along the
Co chains is localized in two orbitals, which are rotated 45 degrees to
each other, having their lobes in the hexagonal plane (d$_{x^2-y^2}$ and 
d$_{xy}$, respectively). In addition we see some spin density on the O atoms
indicating the importance of the 90$^\circ$ superexchange mentioned above.

\subsection{Densities of states (DOS)}

\begin{figure}
\includegraphics{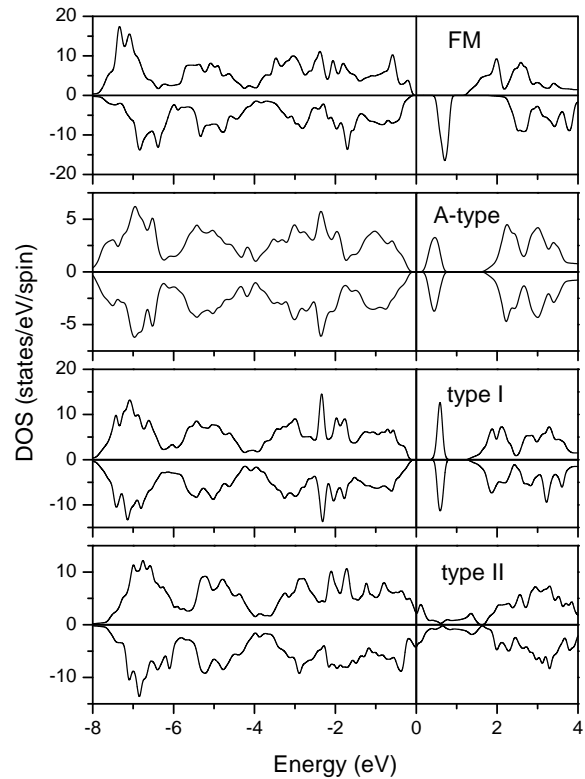}
\caption{\label{dostot}Total density of states (DOS) for the different magnetic
configurations in the alternating orbital ordered state. 
Semiconducting behavior is related to symmetry breaking along the
Co chains. The energy is taken with respect to E$_F$.
}
\end{figure}

In this section we discuss the main features of the
electronic structure of the system using the DOS plots. 
In GGA, the total moment of 2 $\mu_B$ per unit cell is fairly delocalized
over the entire unit cell and thus leads to a relatively small moment
inside the Co spheres (see Table \ref{energytab}).
Within the LDA+U method, we obtain a physically more plausible solution,
since magnetic moments are more localized in the Co
spheres, as expected for a correlated system (see Table \ref{magnldau}). 
Also, we get values of the magnetic moment in agreement with
the experimental LS state and the ``hole'' (see above) is localized in one of
the t$_{2g}$-like orbitals. Using the GGA
approximation we cannot obtain the experimentally found semiconducting
state, not even when we break the symmetry by a Peierls distortion. 
This semiconducting state, however, can be obtained by means
of the LDA+U approach.

In Fig. \ref{dostot} we show the total DOS for the four magnetic
configurations studied. All of them are in an alternating orbital ordered
configuration, since this is their lowest energy state. The main
observation is that insulating behavior is found for the
FM, the A-type AF and AF-type I structures, where alternating orbital
ordering occurs along the Co chains. In the
AF-type II case we forced the translational symmetry along the Co
chains and thus excluded intra-chain orbital ordering 
(we have FM chains coupled
AF to some of their neighbors). In this case, no splitting appears
in the bands around the Fermi level. Hence, the lowest energy states with an
alternating orbital ordering reproduce the experimental evidences, while
without this symmetry breaking a metallic solution would occur.

\begin{figure}
\includegraphics{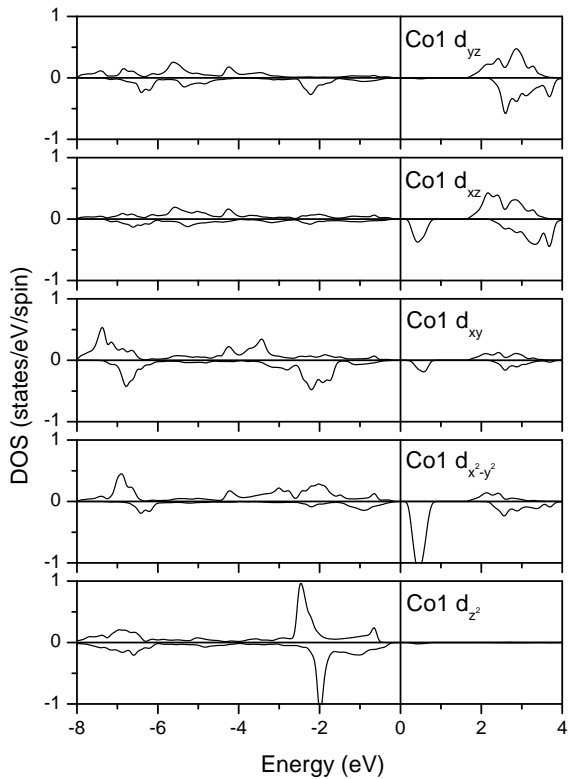}
\caption{\label{dosCo1}Partial Co d DOS for the Co1 atom in a
ferromagnetic, alternating orbital ordered case. The hole is located at the 
d$_{x^2-y^2}$ orbital.
}
\end{figure}

\begin{figure}
\includegraphics{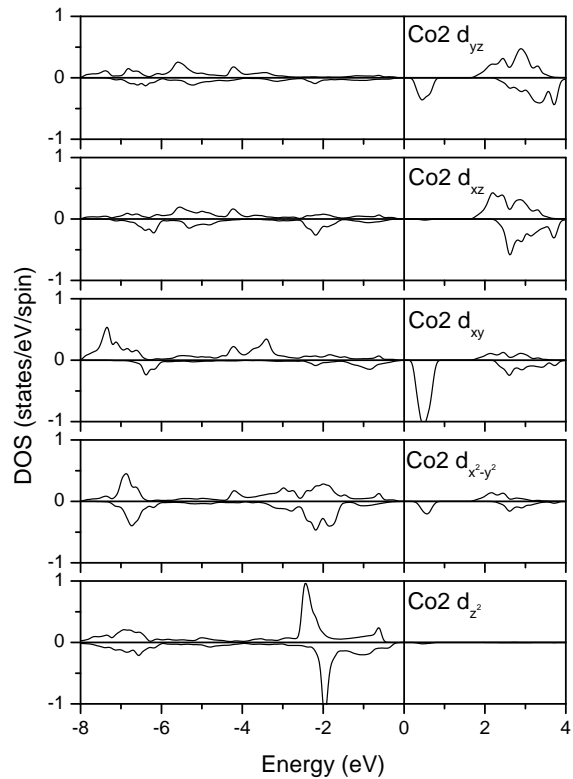}
\caption{\label{dosCo2}Partial Co d DOS for the Co2 atom in a
ferromagnetic, alternating orbital ordered case. The hole is located at
the d$_{xy}$ orbital.
}
\end{figure}

In the case of FM coupling and alternating orbital ordering, 
we show in Figs. \ref{dosCo1} and \ref{dosCo2} the partial DOS corresponding 
to the different
Co d orbitals for the two types of adjacent Co atoms along the Co chains.
For Co1, the hole is mainly located in the  d$_{x^2-y^2}$ orbital with some
d$_{xy}$ and d$_{xz}$ component (see peaks just above E$_F$), where the 
latter produces the tilting visible in the
spin-density shown in Fig. \ref{rho}.
In the case of Co2, the hole is mainly of d$_{xy}$ character with some
admixture of d$_{yz}$ and d$_{x^2-y^2}$. The e$_g$-like orbitals (d$_{xz}$
and d$_{yz}$) are predominantly unoccupied while the d$_{z^2}$ orbital is 
fully occupied in any case.

\begin{figure}
\includegraphics{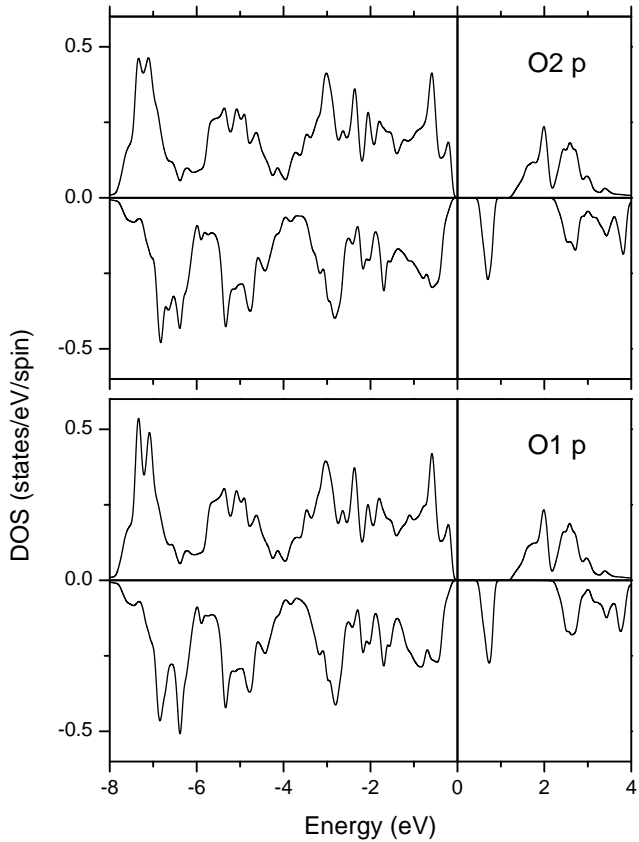}
\caption{\label{dosOp}Partial O p DOS for the ferromagnetic, orbital
ordered case. Observe that the hole has some O p character, which comes
from  p$_x$
and p$_y$ orbitals but has no component along the Co chain axis.
}
\end{figure}

In Fig. \ref{dosOp} we observe that the hole found in the total
DOS plot has some O p character as well. This is again a manifestation of the
importance of the 90$^\circ$ superexchange mechanism. It is mainly due to 
the p$_x$ and
p$_y$ but not due to the p$_z$ orbital which points along the chain axis. One
can see this  from Fig.
\ref{rho} as well, noticing that the O spin cloud is mainly in the plane
perpendicular but not along the Co chains.

\section{Summary}

In this paper we carried out ab initio density functional theory calculations
using the full-potential APW+lo method as implemented in the WIEN2k code
on the transition metal oxide BaCoO$_3$. 

The calculations within GGA to the exchange-correlation
potential gave a ferromagnetic configuration as the most
stable one. The value of the spin magnetic moment was calculated using
the FSM method and the result was 1.5 $\mu_B$/unit cell, smaller than the
expected 2 $\mu_B$/unit cell for a perfect LS state.
Antiferromagnetic configurations within the Co chain could not be obtained. 
All the GGA solutions are metallic in contrast
to the experimental semiconducting state.
We investigated the possibility of a Peierls
distortion as a cause for the semiconducting behavior, but 
had to rule it out by total energy calculations.

Within the LDA+U method, we could incorporate the
correlation effects among the Co 3d electrons. 
We studied different magnetic and orbital orderings and found that
an intra-chain alternating orbital ordering with FM coupling 
is the most stable solution. 
The orbital ordering is caused by a nearly 90$^\circ$ superexchange
mechanism of the face-sharing octahedra and leads to the main
stabilization, whereas the magnetic order is of secondary importance.
In addition, this orbital ordering produces an insulating solution, which
reproduces the experimental observations.

Our electronic structure calculations indicate that the inter-chain
couplings are weak but stronger than expected for a nearly 1D system.

\begin{acknowledgments}

V.P., M.I., D.B. and J.E.A. wish to thank the CESGA (Centro de 
Supercomputaci\'on
de Galicia) for the computing facilities, J. Rivas and J.
Castro for fruitful discussions and also acknowledge
the Xunta de Galicia for the financial support through the
project PGIDIT02TMT20601PR. V. P. wishes to thank the Xunta de Galicia for the
financial support. P.B. acknowledges the support by the Austrian Science Fund
(FWF), Project No. P14699-TPH.
\end{acknowledgments}

\newpage %Just because of unusual number of tables stacked at end
\bibliography{prb}% Produces the bibliography via BibTeX.

\end{document}